\documentclass[prb,aps, reprint, onecolumn nofootinbib, amsmath ,amssymb, superscriptaddress, 10pt, citeautoscript, nobibnotes,]{revtex4-2}

\usepackage{graphicx} 
\usepackage{braket}
\usepackage{quantikz}
\usepackage{comment}
\usepackage{dsfont}
\usepackage[normalem]{ulem}
\usepackage{hyperref} 
 \hypersetup{
    colorlinks=true,
    citecolor=blue,  
    linkcolor=blue,  
    filecolor=magenta,      
    urlcolor=blue,
}

\makeatletter
\def\blfootnote{\xdef\@thefnmark{}\@footnotetext}

\begin{document}
\blfootnote{This manuscript has been authored by UT-Battelle, LLC, under Contract No. DE-AC0500OR22725 with the U.S. Department of Energy. The United States Government retains and the publisher, by accepting the article for publication, acknowledges that the United States Government retains a non-exclusive, paid-up, irrevocable, worldwide license to publish or reproduce the published form of this manuscript, or allow others to do so, for the United States Government purposes. The Department of Energy will provide public access to these results of federally sponsored research in accordance with the DOE Public Access Plan.}

\title{
Towards Predictive Quantum Algorithmic Performance: 
\\Modeling Time-Correlated Noise at Scale}
\author{Amit Jamadagni}  
\email{gangapurama@ornl.gov}
\affiliation{Computational Sciences and Engineering Division, %
Oak Ridge National Laboratory, %
Oak Ridge, Tennessee 37831, USA}
\author{Gregory Quiroz}
\email{gregory.quiroz@jhuapl.edu}
\affiliation{The Johns Hopkins Applied Physics Laboratory, Laurel, Maryland, 20723, USA}
\affiliation{William H. Miller III Department of
Physics and Astronomy, Johns Hopkins University, Baltimore, Maryland 21218, USA}
\author{Eugene Dumitrescu}  
\email{dumitrescuef@ornl.gov}
\affiliation{Computational Sciences and Engineering Division, %
Oak Ridge National Laboratory, %
Oak Ridge, Tennessee 37831, USA}
\date{\today}

\newcommand{\subfig}[2]{%
    {#1} \vtop{
  \vskip0pt
  \hbox{#2}
}}
\begin{abstract}
Combining tensor network techniques with quantum autoregressive moving average models, we quantify the effects of time-correlated noise on quantum algorithms and predict their performance at scale. As a paradigmatic test case, we examine the quantum Fourier transformation. Building on our first technical result, which shows how stochastic tensor network calculations capture frequency correlations, our second result is the revelation that infidelity exponents (scaling from diffuse, to superdiffuse) are determined by the spectral features of the noise. This numerical result rigorously quantifies the common belief that the temporal correlation scale is a key predictive feature of noise's deleterious impact on multi-qubit circuits. To highlight prospects for predicting algorithmic performance, our third result quantifies how infidelity scaling exponents--- which are fits determined by training data at moderate scales (40-80 qubits)---can be used to predict more computationally expensive simulation at larger scales (100-128 qubits). Aside from highlighting the scalability of our methods, this workflow feeds into our last result, which is the proposal of predictive benchmarking protocols connecting simulations to experiments. Our work paves the way for large-scale algorithmic simulations and performance prediction under hardware-relevant noise conditions informed by realistic device characteristics.
\end{abstract}

\maketitle

\section{Introduction}

Classically simulating quantum systems is key to realizing the promise of quantum computing. Simulations verify and predict microscopic hardware dynamics, macroscopic quantum computational models, and are critical tools for achieving error resilience across the quantum stack. For example, simulations are utilized in quantum device design and optimization~\cite{goerz2017charting, gao2021sim, edlbauer2022semiconductor}, the development of low-level quantum control protocols~\cite{peirce1988qoc, leung2017qoc, abdelhafez2019qoc, gunther2021quandary}, and the evaluation and design of quantum error-correcting codes~\cite{Darmawan2017, gottesman1997stabilizer, gottesman1998qec, aaronson2004qec, anders2006qec} and more general quantum algorithms~\cite{haner20175, villalonga2019flexible, guo2019peps, huang2021efficient, kissinger2022simulating, bravyi2022hybrid}. Additionally, the classical simulation of algorithmic performance and resource analyses quantify quantum advantage~\cite{Arute2019,cross2019, kim2023evidence}. 

Despite its importance, the classical simulation of quantum dynamics remains challenging. In general, the computational cost of such simulations scales exponentially with the number of qubits, restricting exact simulations of quantum dynamics to only tens of qubits. Extending beyond this regime typically requires approximation methods or imposing structural constraints on the dynamics, both of which may be physically inaccurate. Furthermore, to model precision errors in intended control operations and also the effects of unwanted system–environment interactions, higher-dimensional dynamical noise maps may be required.

Real hardware is subject to both Markovian and non-Markovian error sources, so modeling both sources is essential. The former lends itself to dissipative dynamics described by the Lindblad master equation (LME)~\cite{lidar2020lecturenotestheoryopen, manzano2020short} or approximations thereof.
While there is subtlety in the definition of non-Markovian noise~\cite{breuer2016nm}, to first order they can be described as including temporally correlated processes. Such processes' deleterious influence is well documented across the quantum stack and hardware platforms, such as semiconductor spin qubits~\cite{chan2018spin-qns}, superconducting circuits~\cite{bylander2011noise, yan2016flux, vonlupke2020qns}, neutral atoms~\cite{burgardt2023measuring, yu2013suppressing, day2022limits, lukashov2025qns, manetsch2025tweezer}, and trapped ions~\cite{frey2017application, milne2021trapped, nakav2023trapped, frey2020qns, wang2017single}. Its distinctive error signatures have been extensively studied in noise characterization protocols, such as randomized benchmarking~\cite{knill2008rb, ball2016rb, Figueroa2021rb, figueroa2022towards, brillant2025rb} and gate set tomography~\cite{greenbaum2015gst, blume2017demonstration, nielsen2021gate, li2024non, vinas2025microscopic}, and have motivated the development of quantum noise spectroscopy~\cite{alvarez2011qns, pazsilva2017qns, szankowski2017qns}. Temporally correlated noise has also been shown to significantly affect the performance of quantum error correction~\cite{clemens2004qec, ng2009qec, aharonov2006qec, clader2021qec} and variational quantum algorithms~\cite{quiroz2021, kattemolle2023vqe}.

However, as the sophistication of the noise model increases, technical and conceptual layers of complexity add up, and the associated computational cost grows accordingly. This necessitates simulation strategies that strike a balance between scalability and the ability to faithfully represent relevant noise processes. Such trade-offs must also be considered when employing classical simulations for verification, prediction, and design tasks.

Tensor network (TN) methods have recently emerged as a leading framework for the classical simulation of unitary and non-unitary quantum dynamics in many-body systems~\cite{Schollwock2011}. The efficiency of TN approaches stems from their power to effectively represent quantum state's entanglement, thus enabling compact representations. Among TN methods, matrix product states (MPS) provide a powerful and well-developed ansatz, where low-entanglement states can be efficiently represented and manipulated~\cite{Hastings_2007}. Entanglement-adaptive scaling allows TN methods to access a substantially broader class of dynamics than alternative approaches, such as Clifford-only simulations, which restrict operations to a non-universal gate set~\cite{Aaronson2004, Bennink2017,Nguyen2022,McCaskey_2018}.  As a result, TN techniques have been widely applied to the study of Hamiltonian dynamics, quantum quenches~\cite{Jamadagni2024_kb}, error correction~\cite{Kobayashi2024}, and even Shor's algorithm~\cite{Dumi, Dang2019}. They also serve as classical benchmarks for experimental demonstrations of quantum advantage~\cite{Zhou2020,King_2025}. Moreover, recent advances have positioned TNs as fruitful candidates for simulating both Markovian~\cite{Verstraete2004markov, Zwolak2004markov, werner2016markov, cui2015markov, Jorgensen2019nm, cheng2021markov, sander2025large} and non-Markovian environments~\cite{prior2010nm, vega2015nm, strathearn2018efficient, fux2023nm}. The latter has been overwhelmingly rooted in explicit representations of quantum bath degrees of freedom.

This work attempts to bypass the need for modeling the quantum environment, while still preserving relevant temporal correlations.
By pairing the stochastic MPS formalism with Schr{\"o}dinger Wave Autoregressive Moving Average (SchWARMA) models~\cite{Schultz2021} as described in the next section, this work bridges the gap between scalable simulation methods and semi-classical noise representations. 
Since SchWARMA model's noise trajectory averaging may be parallelized, we are able to
explore the scalability of this approach through simulations of quantum algorithms at large scales. To draw broad conclusions regarding the interplay of noise and simulation scalability, the quantum Fourier transform (QFT) is employed as a paradigmatic setting for both training and testing predictability. 

This work is organized as follows. Section~\ref{sec:meth} introduces our numerical tools. This includes time-correlated noise models, techniques to scale simulations, and the performance metrics we utilize. Our main findings are described in Sec.~\ref{sec:algs}. This includes our first key finding of a transition from diffuse to super-diffuse infidelity scaling, as a function of the total noise power. Our second key result is the predictive power of empirical fits from a given system size to a much greater scale, in terms of system size and circuit depth. Lastly, to understand the fundamental limitations of predictive quantum simulations, we present simulations at extreme scales. Towards the end, in Sec.~\ref{sec:Dis}, we present open directions that may generalize our work in terms of noise realism, quantum simulations, algorithmic analysis, and hardware benchmarking. 

\section{Methods}
\label{sec:meth}
\subsection{Schr\"{o}dinger-wave autoregressive moving-average models (SchWARMA)}
\label{sec:ARMA}

Autoregressive Moving Average (ARMA) models provide a compact and flexible framework for describing time-series data with temporal correlations. Motivated by the presence of temporally correlated noise in quantum circuits, prior work~\cite{Schultz2021} introduced the SchWARMA framework, which extends classical ARMA models to quantum dynamical maps. By lifting ARMA processes to operate on quantum channels rather than classical signals, SchWARMA enables the systematic modeling of time-correlated noise in both unitary and non-unitary error processes. This approach goes beyond traditional Markovian noise models---such as stochastic Pauli channels and LME descriptions---which typically assume memoryless dynamics. In contrast, SchWARMA captures non-Markovian temporal structure while retaining computational tractability. Moreover, the framework naturally accommodates a broader class of coherent (unitary) error mechanisms, allowing for more realistic representations of hardware noise that exhibits both temporal correlations and coherent structure.

SchWARMA models establish a connection between classical ARMA processes and completely positive trace-preserving (CPTP) maps by parameterizing quantum channels over complex Stiefel manifolds. Consider an ARMA model composed of $L$ independent stochastic processes. For each component $l$, the time-series evolution is given by
\begin{equation}
y_{k}^{(l)} 
= \sum_{i=1}^{p^{(l)}} r_{i}^{(l)} y_{k-i}^{(l)} 
+ \sum_{j=0}^{q^{(l)}} w_{j}^{(l)} x_{k-j}^{(l)},
\end{equation}
where $y_{k}^{(l)}$ denotes the $l^\text{th}$ output at discrete time step $k$, and $x_{k}^{(l)}$ is an input noise process, typically taken to be independent and identically distributed (i.i.d.) Gaussian noise. The coefficients $\{r_{i}^{(l)}\}$ define the autoregressive (AR) component, while $\{w_{j}^{(l)}\}$ define the moving-average (MA) component. The integers $p^{(l)}$ and $q^{(l)}$ denote the AR and MA orders, respectively. When driven by white Gaussian noise, the power spectral density (PSD) of $y^{(l)}$ is given by
\begin{equation}
S_{y}^{(l)}(\omega)=
\frac{\left|
\sum_{k=0}^{q^{(l)}} w_{k}^{(l)} e^{-i k\omega}
\right|^2}
{\left|
1 + \sum_{k=1}^{p^{(l)}} r_{i}^{(l)} e^{-i k\omega}
\right|^2}.
\end{equation}
Thus, the ARMA coefficients directly determine the temporal correlation structure. Moreover, ARMA models can approximate any discrete-time power spectrum to arbitrary accuracy~\cite{ives2010analysis}.

To lift this classical stochastic process to the quantum setting, SchWARMA associates the ARMA outputs with generators in the tangent space of a unitary $U_k$. Given a unitary operation $U_k$, the corresponding CPTP map $\mathcal{S}_k$ at time step $k$ is constructed as
\begin{equation}
\mathcal{S}_{k} = \mathcal{R}\left(\sum_{l=1}^{L} y_{k}^{(l)} X^{(l)}\right),
\end{equation}
where each $X^{(l)}$ is an element of the tangent space at $U_k$, and $\mathcal{R}$ denotes a retraction map that projects the tangent-space element back onto the manifold of physical quantum channels (see App.~\ref{app:riemannian_schwarma} for details). This construction ensures that $\mathcal{S}_k$ remains CPTP at every time step while inheriting the prescribed temporal correlations from the underlying ARMA processes.

In this study, we will leverage SchWARMA to study the impact of time-correlated dephasing on quantum algorithms at scale. The corresponding SchWARMA model can be created by setting the unitary $U_{k} = \mathds{1}$, which results in the tangent space elements $X = -i\sigma_{z}$, where $\sigma_{z}$ is the Pauli-Z operator. This yields the CPTP map
\begin{equation}
\mathcal{S}_{k} = \begin{pmatrix}e^{-iy_{k}} & 0 \\ 0 & e^{iy_{k}}\end{pmatrix},
\end{equation}
which represents a direct connection between the ARMA process and dephasing noise.

\subsection{Time-correlated noise}
We now specify a family of time-correlations for the noise that we express as an ARMA model. Consider the Ornstein-Uhlenbeck (OU) process, described by the differential equation
\begin{equation}
dx_t = \theta(\mu-x_t) dt + \sigma dW_t.
\label{eq:ou_process}
\end{equation}
Here $dW$ denotes a Wiener process, $\sigma$ is the strength of stochastic perturbations, $\theta$ is a non-negative decay rate, and $\mu$ is the long-time mean. In our analysis, it will be helpful to work with the correlation time $\tau = 1/\theta$. The noise spectrum corresponding to the OU process is
\begin{align}
S(\omega) &= \frac{\sigma^2}{\theta^2 + \omega^2}.
\end{align}
This spectrum results in a noise power, $P$, given by
\begin{align}
P &= \int\limits_{-\infty}^{\infty} d\omega {S(\omega)} = \int\limits_{-\infty}^{\infty} d\omega  {\frac{\sigma^2}{\theta^2 + \omega^2}} 
   = \frac{\sigma^{2}\pi}{\theta}
\label{eq:noise_power}
\end{align}
The OU process defines an ARMA process with $(p,q)=(1,0)$, with coefficients $r_1=\Omega$ and $w_0=\frac{\sigma^2}{2\theta}(1-\Omega^2)$, where $\Omega=\exp(-\theta dt)$.

We utilize this model to describe a quantum circuit subject to dephasing noise. That is, at times $t_i$ during a circuit, we subject every qubit, $q$ to a dephasing operation $R_{Z}(y^{(l)}_{(q,i)}) = \exp{-i\frac{y^{(l)}_{(q,i)}}{2} Z}$ (Pauli-$Z$ parameterized by $y^{(l)}_{(q,i)}$), with the $y^{(l)}_{(q,i)}$'s generated by Eq.~\ref{eq:ou_process}. For the sake of brevity, in the rest of the article, we drop the qubit index $q$ while noting that the parameters $y^{(l)}_{i}$ are independently generated for each qubit, $q$. We do not account for distinctions in gate time and thus, all gates are assumed to be equivalent in duration. Idle periods are also appended with an error gate to model qubits awaiting further algorithmic operations. An illustration of an instance of a SchWARMA-fied noisy circuit trajectory is given in Fig.~\ref{fig:qft_noise}. We employ the Mezze package~\cite{Murphy2022} to generate the corresponding noisy $\{y^{(l)}_i\}$. Written in units of the gate time, $t_g = 100ns$ being roughly the geometric average of superconducting one- and two-qubit gates, the correlation time is given by $\tau = \alpha t_{g}$, with the single-qubit instantaneous noise power being $P_{1Q} = \sigma^{2} \pi \alpha t_{g}$. Given a system of $N$ qubits, each independently undergoing identical OU processes, the total noise power is 
\begin{equation}\label{eq:total_pow}
P_{tot} = NP_{1Q}.
\end{equation}

Given a quantum circuit of depth $D$, the time-integrated power, i.e., the work done on the system by noise, is $W = \int_0^D dt P_{tot} = D N \sigma^{2} \pi \alpha t_{g}$. 
A quantum algorithm's depth typically scales as some low order polynomial in the system size. In the next section we consider an algorithm where $D = \mathcal{O}(N^2)$. We will then examine the infidelity scaling with respect to the noise power and circuit depth.

\subsection{Stochastic MPS Formalism}
The computational cost of TN simulations scales as a polynomial in the bond dimension $\chi$ of the network, with typical costs of the form $O(\chi^{k})$, where the exponent $k$ depends on the network geometry and the specific algorithm employed. For MPS-based time evolution, such as time-evolving block decimation (TEBD) and time-dependent density matrix renormalization group (tDMRG), the cost typically scales as $O(\chi^{3})$ per time step~\cite{Vidal2003a}. Alternatively, the approximations in the bond dimension $\chi$ can also be achieved by controlling the approximation error $\epsilon$, with the growth in bond dimension determined by the evolving state's entanglement entropy. In the simulations presented here, we set the truncation cutoff to be on the order of $\epsilon=10^{-14}.$

Stochastic formalisms~\cite{Breuer1999,Moodley2009,Gambetta2002,Lin2020}, ranging from Markovian dynamics, using the Monte Carlo wavefunction method~\cite{Molmer1993}, to recent variants, involving tensor networks to sample mixed quantum states~\cite{Cichy2025}, are useful in scaling simulations. SchWARMA models simulate the correlated noise dynamics by interspersing the noiseless quantum circuits with noisy unitaries that are sampled from an underlying ARMA model. To analyze the effect of noise on a given quantum circuit at scale, we employ the MPS toolkit~\cite{Schollwock2011}. That is, we prepare the noisy/noiseless quantum states by contracting the circuit and noise unitaries into an input MPS. Furthermore, due to the stochastic nature of the SchWARMA models, we average relevant properties over many noisy unitary MPS trajectories, denoted by $n_{t}$ trajectories, resulting in the stochastic MPS formalism. 

With SchWARMA models being stochastic in nature and MPSs allowing for analysis at scale, combining these tools results in a powerful framework for simulating noisy quantum systems along both scale and noise-complexity axes. One additional computational advantage, and feature of the SchWARMA-MPS formalism, is that the trajectories are independent, which allows for massive parallelization at scale.

\subsection{Noise Metrics}
Metrics are required to quantify the effect of noise. Various metrics could in principle be considered, each providing different information regarding average or worst-case performance, with respect to states or operators. For simplicity, this work makes use of the (in)fidelity of quantum states. Typically, these results are state dependent, but can be representative of average use cases and provide bounds on operator metrics. If desired, the fidelity can also be related to trace distances~\cite{Jozsa1994,Lidar2008}. 

Denote the ideal quantum state as $\ket{\psi} = U \ket{\eta}$ where $\ket{\eta}$ represents an input state. For the moment, $U$ represents a quantum algorithm which has been abstracted away.  Since this is a pure state, expressions for fidelity between $\ket{\psi}$ (as a density operator $\rho_{i} = \ket{\psi}\bra{\psi}$) and other states will simplify. Subjected to temporally correlated noise across many trajectories results in the averaged statistical ensemble $\rho_{n} = \frac{1}{n_{t}}\sum_{l=1}^{n_{t}} \tilde{U}_l |\eta\rangle \langle \eta |\tilde{U}_l^\dagger$, where $\tilde{U}_{l}$ is $l^\text{th}$-realization of noisy unitary among $n_{t}$ trajectories. Since the target state is pure, the fidelity with the noisy state is
\begin{eqnarray}
    \mathcal{F}(\rho_{i}, \rho_{n}) &=& \text{Tr}[\rho_{i} \rho_{n}] = \braket{\psi| \rho_{n} |\psi} \\
    &=& \frac{1}{n_{t}}\sum_{l=1}^{n_{t}}  \langle  \eta  |U^\dagger \tilde{U}_l | \eta \rangle \langle \eta  |\tilde{U}_l^\dagger U | \eta  \rangle \nonumber \\
    &\mathcal{F}_\eta=& \frac{1}{n_{t}} \sum_{l=1}^{n_{t}}  |\langle  \eta  |U^\dagger \tilde{U}_l | \eta \rangle|^2 
\end{eqnarray}
This shows us that the fidelity with respect to an input state $\ket{\eta}$ ($\mathcal{F}_\eta$) is simply the average over the individual trajectory's fidelities. Approaching zero when errors vanish and unity when the algorithm is completely incorrect, we use infidelity 
\begin{equation}
\label{eq:infi}
\mathcal{I} = 1-\mathcal{F}
\end{equation}
as a metric to quantify algorithmic errors. We can now relate this expression to prior algorithmic error bounds. Later, in Sec.~\ref{sec:scale}, we will revisit this quantity and apply it to validate our numerical results at scale. 

\begin{figure}[t!]
\begin{center}
\includegraphics[width=\linewidth]{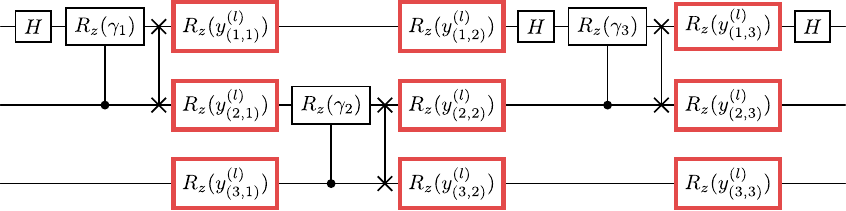}
\end{center}
\caption{Graphical circuit representation of an $N=3$ qubit noisy QFT. The QFT circuit (black) is deformed with time-correlated dephasing noise (red gates). The $y^{(l)}_{(q,i)}$s ($q^\text{th}$ qubit, $i^{\text{th}}$ time step and $l^{\text{th}}$ instance of the ARMA model) are sampled from a physically relevant spectrum via the SchWARMA formalism. The $y^{(l)}_{i}$s on the same qubit at different time steps are correlated (time correlated) while $y^{(l)}_{i}$s on different qubits at the same time step remain uncorrelated (spatially uncorrelated) and are sampled from independent SchWARMA processes.}
\label{fig:qft_noise}
\end{figure}

\section{Algorithmic Applications}
\label{sec:algs}

\subsection{The Quantum Fourier Transform}
\label{sec:QFT}

By exponentially reducing the computational complexity, compared to the conventional and fast Fourier transformations, the QFT has become a key subroutine in many important quantum algorithms, for example, quantum phase estimation~\cite{kitaev1995qpe} and Shor's algorithm~\cite{shor1999polynomial}.  Analyzing the impact of correlated noise in this key subroutine will therefore provide broad algorithmic insights into practical implementations on current quantum architectures. 

In an $M=2^{N}$ dimensional Hilbert space, the QFT transforms an $N$-qubit quantum state, $\ket{\psi} = \sum\limits_{n=0}^{M-1}a_{n}\ket{n}$, as
\begin{equation}
\text{QFT}\ket{\psi} = \sum_{n=0}^{M-1}\tilde{a}_{n}\ket{n},
\end{equation}
where $\tilde{a}_{n} = \frac{1}{\sqrt{M}}\sum\limits_{j=0}^{M-1}a_{j}e^{\frac{2i\pi n}{M}}$. At the circuit level, the QFT is defined in terms of long-range $CR_Z(\gamma)$ (controlled Pauli-$Z$ parameterized by $\gamma$) gates.  While there exist multiple strategies to simulate the QFT circuit, we consider the variant where a SWAP gate is included after each $CR_Z(\gamma)$ gate, resulting in a two-local circuit as illustrated in Fig.~\ref{fig:qft_noise}. This choice results in a circuit that is amenable to simulation by the MPS data-structure~\cite{Woolfe2014,Fowler2004}. In either case, the $N$-qubit QFT gate complexity scales as $\mathcal{O}(N^2)$. 
 
To model time-correlated noise's impact on the QFT, we interleave the ideal circuit with dephasing $R_{Z}(y^{(l)}_{i})$ gates, where $y^{(l)}_{i}$'s are sampled from the corresponding ARMA model, here the OU process, as described in Sec.~\ref{sec:ARMA}. While there exist many choices for how to inject dephasing noise, given that we fix the gate times to 100ns, we induce them after each SWAP gate, see Fig.~\ref{fig:qft_noise}. For a $N$-qubit QFT, the number of injected noise gates scale as $\mathcal{O}(N^3)$. 

\subsection{Error Scaling}
\label{sec:prior}
In many areas of physics, ballistic versus diffusive scaling describes how quantities spread with time. If a particle, wave packet, or excitation retains a well-defined velocity, its displacement grows linearly $\sqrt{\langle\Delta x^2\rangle} \propto t$. On the other hand, repeated randomizing collisions lead to diffusion with $\sqrt{\langle\Delta x^2\rangle} \propto \sqrt{t}$. A textbook example is Brownian motion~\cite{Einstein1905}, which crosses over from short-time inertial/ballistic behavior to long-time diffusion and has been directly observed in experiments resolving the instantaneous-velocity regime~\cite{Li2010}. In condensed-matter physics, phonon-mediated heat flow becomes ballistic (or quasi-ballistic) when device dimensions approach phonon mean free paths~\cite{Anufriev2021}, contrasting with ordinary diffusive conduction in bulk. In interacting many-body systems, one can even see ballistic entanglement growth coexisting with diffusive transport of conserved quantities, emphasizing that different spreads in the same quantum system can fall into different scaling classes~\cite{Kim2013}. 

In quantum computing, coherent quantum-walk dynamics exhibit ballistic wave packet spreading, while decoherence/disorder can drive a transition to diffusive classical-walk scaling~\cite{Yin2008, Schreiber2011}. Further, the linear addition of coherent errors and diffuse spread due to randomized trajectories is the rationale underlying powerful randomized compilation techniques~\cite{Wallman2016,Flammia2020}.  Here, by averaging over sufficiently many randomization realizations at the compilation level, the effects of coherent noise (where $\mu \gg \sigma$) are quadratically reduced by transforming errors into effectively stochastic cases (where $\mu \ll \sigma$). 

Previous works have studied the scaling of errors within specific algorithms. For example, in the context of precision noise in quantum approximate optimization algorithms~\cite{quiroz2021}. There, noisy unitaries $\tilde{U}_l$ were constructed by shifting the optimal parameters, of the ideal unitaries, with noise that is sampled from a normal distribution $\mathcal{N}(\mu,\sigma^2)$ with mean, $\mu$, and variance, $\sigma^2$. Specifically, two different regimes were analyzed. One regime, $\mu=0, \sigma \neq 0$, gave rise to stochastic errors and displayed diffuse error scaling. Another regime, namely $\mu \neq 0, \sigma = 0$, generated coherent errors which led to ballistic algorithmic error~\footnote{$\mu$ was kept constant across trajectories}. Denoting the multi-trajectory average $\overline{\tilde{U}} = \frac{1}{n_{t}}\sum_{l=1}^{n_{t}} \tilde{U}_l$, they found the following scaling relationships with respect to circuit depth:
\begin{equation}
\label{eq:ave_unitary_error}
|| U - \overline{\tilde{U}}||_{2} \propto
\begin{cases}
 \sqrt{P_{tot}D}; \hspace{0.2cm} \text{stochastic errors} \\
P_{tot}D; \hspace{0.42cm} \text{coherent errors} \\
\end{cases}
\end{equation}
where $\|A\|_2=\sqrt{\text{Tr}(A^\dagger A)}$ denotes the Frobenius norm.

\subsection{Numerical Results}
In this section, we analyze the scaling of the infidelity, $\mathcal{I}$, as a function of total noise power and circuit depth in various regimes that parameterize OU processes. To this extent, we analyze the infidelity in the regimes of $\alpha \gg \sigma$ and $\sigma \gg \alpha$. In addition, motivated by the results in Ref.~\cite{quiroz2021}, we also extract the scaling as a function of the noise power and circuit depth.

We first hypothesize the scaling behavior that should occur in the different regimes. Afterwards, we will present empirical results using numerical simulations. In the first case of $\alpha \gg \sigma$, i.e., in the regime where correlation times are the dominant parameter, the $y^{(l)}_{i}$ rotations coherently accumulate over a long period of time. However, this coherent accumulation is tempered by i) the fact that different trajectories are independent of one another and ii) the fact that, in the Heisenberg picture, the noise itself is transformed by the unitary circuit. In the second case, where $\sigma \gg \alpha$, stochastic perturbations kick the $y^{(l)}_{i}$ parameters randomly, with almost no correlations from one time to the next. This leads to rotations that, when taken as a whole, exhibit Brownian motion. Therefore, for a given noise power, in the case of $\alpha \gg \sigma$, the slow variation in $y^{(l)}_i$ leads to a larger, more coherent drift in $\tilde{U}$ compared to the case of $\sigma \gg \alpha$. In addition, it is trivial to observe that increasing the total noise power will increase the infidelity.

To test the above hypothesis, we employ the stochastic MPS formalism and compute the infidelity in each regime. For each value of $\sigma$ and $\alpha$, we sample the perturbed angles $\{y^{(l)}_{i}\}$ from the corresponding OU process.
We then simulate the QFT circuit, interspersed with the noisy gates over $n_{t}$ independent noisy trajectories, and compute the corresponding infidelity, $\mathcal{I}$, as in Eq.~\eqref{eq:infi}. For now, note that the input state $\ket{\eta}$ on which the QFT and noisy QFT act upon is a entangled random MPS characterized by maximum bond dimension $\chi=4$. We use ITensors.jl~\cite{ITensors2022} to simulate the action of unitaries on the MPSs.

\begin{figure}
\begin{center}
\includegraphics[width=\linewidth]{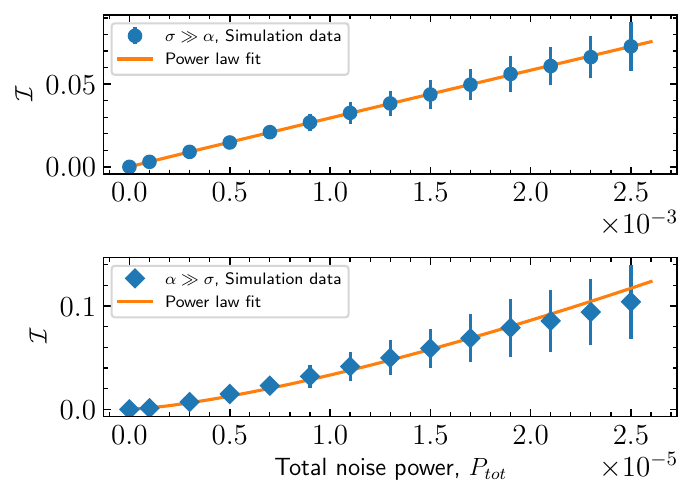}
\end{center}
\caption{Infidelity, $\mathcal{I}$ as a function of total noise power, $P_{tot}$ for the different regimes of (top) $\sigma \gg \alpha$, (bottom) $\alpha \gg \sigma$, for a system size of $N=40$ qubits. For the input state, $\ket{\eta}$, we consider a random MPS characterized by maximum bond dimension, $\chi=4$. In both regimes, we fit the  $\mathcal{I}$, to a power law scaling i.e., $\mathcal{I} = \lambda P_{tot}^{\xi}$ where the exponents are given by (top) $\xi = 0.98 \pm 0.06$ (bottom), $\xi = 1.38 \pm 0.08$.} 
\label{fig:qft_noisy_res_1}
\end{figure}

To systematically investigate the scaling relations in the different regimes, we employ Eq.~\eqref{eq:total_pow} to consistently vary $\alpha$ and $\sigma$, respectively. In the $\alpha \gg \sigma$ case, we first fix $\sigma=\sigma_{c}$ to a constant value. For a desired power range, say $[P_{tot}^{0}, P_{tot}^{1}]$, we obtain $\alpha_{i} \in [\alpha_{0}, \alpha_{1}]$ such that $\alpha_{i} = \frac{P^{i}_{tot}}{\pi t_{g} N \sigma_{c}^{2}}$ for $P_{i} \in [P_{tot}^{0}, P_{tot}^{1}]$ . Similarly, for the case where $\sigma \gg \alpha$, we first fix the total noise power range to $[P_{tot}^{0}, P_{tot}^{1}]$ and choose a constant $\alpha = \alpha_{c}$. This results in a range of values $\sigma \in [\sigma_{0}, \sigma_{1}]$ where $\sigma_{i} = \sqrt{\frac{P_{tot}^{i}}{\pi t_{g} N \alpha_{c}}}$. Having obtained the values for $\alpha$ and $\sigma$ that dictate the OU process, we generate the correlated angles $y^{(l)}_{i}$ that parameterize the time correlated dephasing noise.

In Fig.~\ref{fig:qft_noisy_res_1}, we present the numerical scaling relations in the $\sigma \gg \alpha$ (circle markers) and $\alpha \gg \sigma$ (diamond markers) regimes. The different markers indicate the infidelity, $\mathcal{I}$, with the errors given by the standard deviation in infidelity computed over $n_{t} = 1000$ noise trajectories. We perform a power-law fit to the infidelity, as a function of total noise power, given by $I = \lambda P_{tot}^{\xi}$ and interpret the results as follows.

\begin{figure*}[t!]
\begin{center}
\begin{tabular}{cp{0.01mm}c}
\subfig{(a)}{\includegraphics[width=0.45\linewidth]{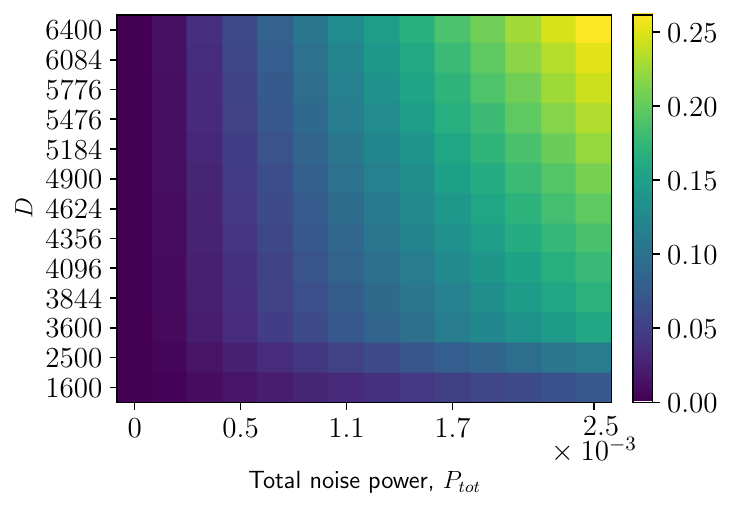}} 
&&
\subfig{(b)}{\includegraphics[width=0.425\linewidth]{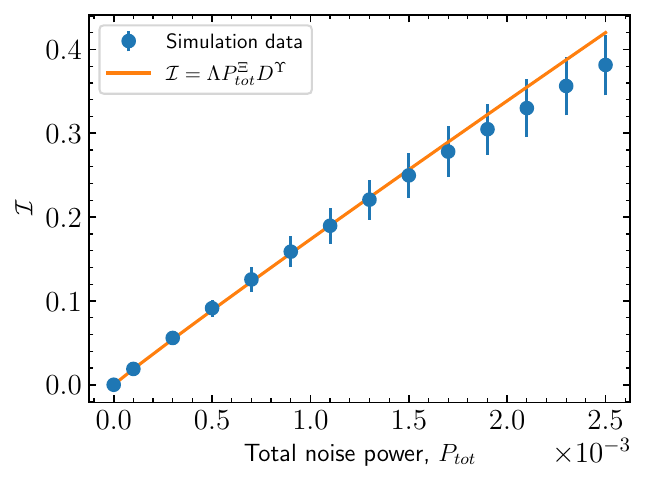}} 
\end{tabular}
\end{center}
\caption{(a) $\mathcal{I}$ as a function of total noise power, $P_{tot}$ and depth, $D$ in the regime of $\sigma \gg \alpha$. For system sizes ($40 \le N \le 80$), we set the input state, $\ket{\eta}$, to be a random MPS characterized by maximum bond dimension, $\chi=4$. Following the earlier results, to extract the scaling behavior, we perform a power-law fit of infidelity $\mathcal{I}$ as a function of the $P_{tot}$ and $D$, with the fit function given by $\mathcal{I} = \Lambda P_{tot}^{\Xi}D^{\Upsilon}$, where $\Xi = 0.965 \pm 0.013, \Upsilon = 0.958 \pm 0.035$ are the fit parameters for the exponents. (b) The power law fit is used to predict the infidelity, $\mathcal{I}$ (denoted by the orange line) for a larger system size of $N=100$ qubits. The fit captures the trend in the simulation data (denoted by the blue dots) particularly well for low total noise power.}
\label{fig:qft_noisy_res_2}
\end{figure*}

The noise in the $\sigma \gg \alpha$ case is consistent with diffusive scaling~\cite{Duplat2013} while the noise in the $\sigma \ll \alpha$ case is sub-ballistic. Indeed, the $\sigma \gg \alpha$ case is close to sampling from $N(\mu=0,\sigma^2)$, and is consistent with an expected exponent $\xi \approx 1$. On the other hand, the $\alpha \gg \sigma$ case, is less coherent, and thus scales sub-ballistically than sampling from perfectly additive noise from $N(\mu,\sigma^2=0)$. This is because the OU process causes the $y^{(l)}_i$ angles to slowly drift both within a single trajectory and also across different trajectories. As a result, similar to randomized compiling, averaging across different trajectories reduces the average noise, resulting in $1<\xi<2$. This intuition is supported by the increased variation in the fidelity in the distinct trajectories. Further, to establish a relationship with previous results as in Eq.~\eqref{eq:ave_unitary_error}, we note that the average transition amplitude, $|\langle  \eta  |U^\dagger \bar{\tilde{U}}_l | \eta\rangle|$, provides a lower bound on the average unitary operator error, see App.~\ref{app:infidelity_to_delta} for more details. Since the transition amplitude is upper bounded by the square root of the fidelity, we intuitively expect the infidelity to scale roughly as $\delta^2$, where $\delta$ is the polynomial scaling in Eq.~\eqref{eq:ave_unitary_error}. In addition, App.~\ref{app:infidelity_to_delta} also presents an upper bound on the infidelity in terms of $\delta$. In terms of our analogy with diffuse and ballistic growth, our numerical results present the infidelity (scaling as $\langle\Delta x^2\rangle$) rather than the average transition amplitude (scaling as $\sqrt{\langle\Delta x^2\rangle}$).

To complete our empirical scaling analysis, we must also discuss the scaling of infidelity as a function of time, which in this scenario is equivalent to the circuit depth $D$. To extract the scaling relationship, we perform the simulations outlined above for different system sizes. For instance, in the regime of $\sigma \gg \alpha$, for a given power range $[P_{tot}^{0}, P_{tot}^{1}]$ and each $N$, we fix $\alpha = \alpha_{c}^{N}$ resulting in $\sigma_{i}^{N} \in [\sigma_{0}^{N}, \sigma_{1}^{N}]$ using Eq.~\eqref{eq:total_pow}. The $\alpha_{c}^{N}$ and $\sigma_{i}^{N}$ are then used to generate the corresponding $y^{(l)}_{i}$ that characterize the time correlated noisy process. In Fig.~\ref{fig:qft_noisy_res_2}(a), we compute the infidelity as a function of both 
$P_{tot}$ and $D$, which in the case of QFT scales as $\mathcal{O}(N^2)$. We perform a two-dimensional power law fit given by $I = \Lambda P_{tot}^{\Xi}D^{\Upsilon}$ to obtain the scaling relationship of infidelity.
We note that the exponent $\Xi$ obtained from the two-dimensional fit is in good agreement with the exponent $\xi$ obtained earlier for a fixed depth $D$. 

To test the power-law scaling's predictive power, we compute the infidelity as a function of total noise power for a larger system. Figure~\ref{fig:qft_noisy_res_2}(b) shows that the empirical power-law scaling is in good agreement with the data obtained from the simulations at the larger system size. We note that the predictive capabilities remain accurate up to a threshold noise power as the power law scaling softens due to infidelity being bounded from above with a supremum of 1. That is, at higher noise power we expect deviations in the power law scaling, thereby requiring additional correction terms to accurately predict the infidelity. We leave such analysis for future works. The scaling laws we obtain are powerful tools, to not only analyze the performance of quantum algorithms in the presence of time correlated noise, but also are useful resources to benchmark noisy quantum hardware architectures. In the following section, we examine how the methods introduced can be applied to predict a noisy quantum system's performance at future scales.

\subsection{Time Correlated Noise At Scale}
\label{sec:scale}
With the increased number of noisy qubits in hardware architectures, there is a greater need for benchmarking protocols that scale and at the same time remain classically tractable. Our time-correlated noise models, introduced in prior sections, are a step towards realizing benchmarking protocols that satisfy the above criteria. 

In this section, we present a simple protocol that captures the effect of time correlated noise. Motivated by previous profiling protocols, such as randomized benchmarking, we use the return probability after the application of a non-trivial identity circuit. In this case, consisting of a unitary and its inverse, acting on an input product state, as a quantitative measure that captures the impact of noise. That is, given a product state, say $\ket{\nu}$, we evolve it by a unitary, $U$, and its inverse , $U^{\dagger}$ with the corresponding return probability given by $|\braket{\nu|U^{\dagger}U|\nu}|^2$. In the case that the inverse is noise-free the return probability is unity. However, when the inverse is a noisy unitary, $\tilde{U}^\dagger$, the quantity $|\braket{\nu|\tilde{U}^{\dagger}U|\nu}|^2$ captures the impact of noise. 

In the current scenario, we set the unitary $U = \text{QFT}(N)$ to be QFT circuit acting on $N$ qubits with the inverse $\widetilde{\text{QFT}}^\dagger(N)$ representing the time correlated noisy inverse QFT as in Fig.~\ref{fig:qft_noise}. The predictive analysis from the previous section acts as a powerful heuristic to deduce a noise power strength for a given infidelity. However, instead of using the exponents ($\Xi, \Upsilon$) from Fig.~\ref{fig:qft_noisy_res_2}, we recompute the exponents using a random product state as the initial state resulting in ($\tilde{\Xi}, \tilde{\Upsilon}$) = 
$(0.975 \pm 0.014, 0.954 \pm 0.039$). The above values are in close agreement with the values obtained using an initial state with a random MPS with bond-dimension $\chi=4$ as in Fig.~\ref{fig:qft_noisy_res_2}. This establishes the fact that the exponents remain largely independent of the initial state. Given the above exponents, we estimate the total noise power required to maintain a fidelity of around 85\% and label it as $P_{0}$.  We then compute the return probability (from sampling protocol as described below) given the above noise power, $P_{0}$, and note that it is in the expected bounds for infidelity; see Tab.~\ref{tab:infid}.

The probability of returning to the initial state after the application of the noisy unitary deviates from unity i.e., due to the action of the time correlated noise we observe a leakage in the return probability to the initial state. This leads to other bitstrings gaining a non-zero probability. MPSs can be used to compute the probabilities of different bitstrings. However, due to the exponential number of basis states, in order to observe the leakage probability, we resort to sampling over the subset of high-probability bitstrings. To this extent, we sample each of the $n_{t}$ independent noisy MPSs in the $\sigma_{z}$-basis. Here, global bitstrings are sampled from a series of marginal probabilities sweeping across the MPS~\cite{Han2018}. 

\begin{figure}[t!]
\begin{center}
\includegraphics[width=\linewidth]{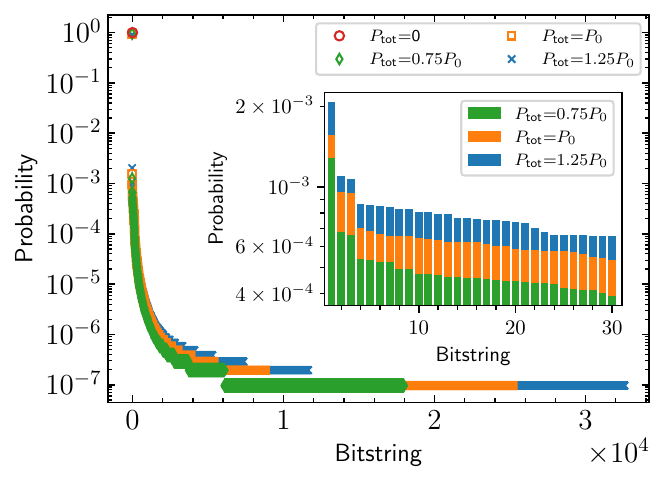}
\end{center}
\caption{Probability of bistrings sampled from a ensemble of noisy states obtained by the application of $\widetilde{\text{QFT}}(N) \text{QFT}(N)$ on a $N=128$-qubit product state for different total noise power, $P_{tot}$ injected into the circuit. We estimate the total noise power $P_{tot} = P_{0}$, using the power law scaling, $\mathcal{I} = \tilde\Lambda P_{0}^{\tilde\Xi}D^{\tilde\Upsilon}$ by fixing the infidelity, $\mathcal{I}$ to be 15\%. The number of bitstring samples is set to $10^4$ for each independent noisy trajectory, with the total number of noisy trajectories set to $10^3$ resulting in a total number of $10^7$ bitstring samples. (Inset) A magnified view of the first 30 high probability bit strings excluding the probability of the initial state. With the increase in power, the return probability to the initial state ($1 - \mathcal{F}$) reduces, with the leakage in probability extending to other bitstrings. At higher powers, the leakage in return probability remains higher in comparison to lower powers thereby reflecting the behavior of the corresponding infidelity.}
\label{fig:qft_iqft_freq_spread}
\end{figure}

In Fig.~\ref{fig:qft_iqft_freq_spread}, we consider the noise regime of $\sigma \gg \alpha$ and present the impact of time-correlated noise on a $N=128$ QFT subroutine. In addition to computing the (approximate) return probability from sampling at a given noise power $P_{0}$, we also compute the leakage of probability into other bit strings at other noise powers and notice that the infidelities are in within acceptable bounds of the expected values shown in Tab.~\ref{tab:infid}. We observe that with increase in the total noise power the return probability further deviates from unity, resulting in a higher infidelity. The drop in the return probability leads to other bitstrings (product states in the $\{\ket{0}, \ket{1}\}^{\otimes N}$ basis) acquiring non-zero probability. 

We briefly summarize the computational resources used to perform the simulations outlined in the article.  The noise trajectories being independent instances, allow for the computational workload to be distributed leading to parallelization. While there exists many strategies to parallelize, we evolve 1000 independent trajectories on 100 cores distributed across 2 computational nodes (50 cores on each node). We note that the cumulative wallclock time for the infidelity computation at a given power instance depends on the entanglement in the initial state. For an initial state characterized by bond dimension $\chi=4$ and a system size of $N=100$, we report a cumulative wall clock time of about $\approx$ 8 hours for one instance of total noise power. However, the cumulative wallclock time considerably decreases to about $\approx$ 1 hour when the input state is a product state. 

\begin{center}
\begin{table}
\begin{tabular}{|p{0.25\linewidth}|p{0.225\linewidth}|p{0.225\linewidth}|p{0.225\linewidth}|}
 \hline
\centering \centering Power, $P$ & \centering $\mathcal{I}_{-} = \tilde{\Lambda}_{-}P^{\tilde{\Xi}_{-}}D^{\tilde{\Upsilon}_{-}}$ & \centering \centering $\mathcal{I} = \newline \tilde{\Lambda}P^{\tilde{\Xi}}D^{\tilde{\Upsilon}}$  & \centering $\mathcal{I}_{s}$ \tabularnewline \hline
\centering 0.75$P_{0}$ & \centering 0.057 & \centering 0.113 & \centering 0.057 \tabularnewline \hline
\centering $P_{0}$ & \centering 0.075 & \centering 0.15 & \centering 0.076 \tabularnewline \hline
\centering 1.25$P_{0}$ & \centering 0.093 & \centering 0.186 &  \centering 0.094 \tabularnewline \hline
\end{tabular}
\caption{Infidelity bounds obtained for a fixed depth of $D=128^2$ and few different scalings of the total noise power, $P_{0}$. The lower bounds for the fit parameters $(\tilde{\Lambda}, \tilde{\Xi}, \tilde{\Upsilon})$ given by $(\tilde{\Lambda}_{-}, \tilde{\Xi}_{-}, \tilde{\Upsilon}_{-})$ correspond to the fit errors obtained by fitting infidelity, $\mathcal{I}$ as a function of total noise power $P_{tot}$ and depth, $D=N^2$. We estimate the fit parameters and corresponding errors by considering $P_{tot} \in [0, 2.5 \times 10^{-3}]$, $N \in [40, 80]$ and using random product states as the initial states for each $N$. $\mathcal{I}$ and $\mathcal{I}_{-}$ correspond to the estimates of infidelity and its lower bound, obtained using the fit parameters, while $\mathcal{I}_{s}$ corresponds to the infidelity obtained from the sampling protocol.}
\label{tab:infid}
\end{table}
\end{center}

\section{Discussion}
\label{sec:Dis}
The steady increase in the number of noisy qubits points to the need for new benchmarking protocols that scale. To this end, the quantum computing community will require advanced simulation techniques to play a larger role in profiling future quantum hardware. Simulating time correlated noise at scale, as demonstrated in Sec.~\ref{sec:algs}, is an important initial step in this direction and provides new pathways to more accurately profile noisy quantum hardware. To analyze the performance of a given hardware device, the simulation technique should be tested against experimentally available quantities. For instance, the probability distribution of sampled bitstrings, or the probability of any given bitstring, can be used as a metric to benchmark hardware performance. In addition, the scaling laws, as a function of the total noise power induced into the circuit, provide new  capabilities to predict and test the performance of architectures that might be beyond the classically simulable regime. 

In the above analysis, to showcase the utility of tensor networks coupled with SchWARMA models, we have restricted our simulations to a subset of all physical quantum noise processes. Namely, the case where noise channels can be described by the collective action of independent trajectories comprising time correlated unitaries. However, SchWARMA models capture a wider class of noisy phenomena that involve time correlated noise channels. Recently, there has been significant progress in representing noisy/mixed quantum states at scale using other tensor network methods, for instance locally purified density operators~\cite{Werner2016, Jamadagni2025_gf}, tree tensor operators~\cite{Wanisch2026} among others. Moving forward, to characterize a wider spectrum of noisy processes, tensor networks representing mixed states can be coupled with time-correlated quantum channels, leading to a more realistic analysis of the impact of the noise on different quantum algorithms. Taken together with our current simulations, this would generalize the types of noise which can be benchmarked, e.g., extending to non-stationary or non-Gaussian noise processes. In addition, it remains to be explored whether the collective action of time correlated unitaries can be mapped to represent a single noise channel, i.e., whether a single quantum channel is sufficient to represent the time-correlated noise. In the case such a mapping exists, it would not only speed up the benchmarking protocols but also potentially aid in real-time characterization of quantum devices. Furthermore, to make the noise characterization even more accurate and realistic, in conjunction with time-correlated noisy unitaries, one might also incorporate spatial correlations. In the context of tensor networks, this could extend to the application of non-local unitaries or CPTP maps, albeit with an increase in the computational requirements. 
 
TNs have also been successfully applied to the simulation of non-Markovian quantum dynamics. A prominent example is the time-evolving matrix product operator method, which employs augmented density tensors to encode the system’s evolution over a finite environmental memory time, together with MPS compression techniques to enhance computational efficiency~\cite{Strathearn2018}. Recently, process tensor formalism has also been employed to study non-Markovian dynamics~\cite{Keeling2025}. It will be of interest to extend our scaling analysis, predictive modeling, and benchmarking to the general setting of non-Markovian dynamics.

Coupled with the fact that quantum computers are inherently noisy, these efforts make the time-correlated noise simulations perfect candidates for both benchmarking and for accurately demarcating the regime of quantum advantage. In most of the experiments that claim quantum advantage, the protocols generally compare against simulation strategies that remain largely agnostic to decoherence. The regimes that remain within the bounds of classical simulability can be used for benchmarking, as outlined above. On the other hand, the regimes that lie beyond the bounds truly signify the regime of quantum computational advantage~\cite{Zhou2020}. Recently, there have been efforts in efficiently representing highly mixed quantum states using tensor networks~\cite{Jamadagni2025}. By removing  spurious correlations from mixed-state tensor network representations, these works address the important nuance of efficiently representing mixed states.

\section{Conclusions}

To summarize, we have studied the effects of time-correlated noise on quantum algorithms at scale. To achieve this, we have combined tensor network techniques with SchWARMA models, the quantum version of time series ARMA models. We demonstrate simulation at scale by considering time-correlated dephasing noise governed by an OU process. We then employ the stochastic MPS formalism, wherein we evolve an initial state with a quantum circuit interspersed with noisy unitaries sampled from a SchWARMA model representing the OU process.

To illustrate the power of the simulation technique, we consider the quantum Fourier transform subroutine in the presence of time-correlated noise. Specifically, we analyze the scaling behavior of the infidelity in the different regimes of the OU process as a function of the total noise power induced into the circuit. Our numerical results reinforce the common folklore that the scaling of the temporal correlations remain a key predictive feature of noise's deleterious algorithmic impact. In doing so, our numerical calculations reveal a crossover from diffusive to sub-ballistic scaling in the algorithmic error. 

By estimating the infidelity at larger system sizes, we also examine the predictive power of our simulations. We observe that the estimated infidelities from the scaling training data are in good agreement with larger-scale simulation results. Motivated by this, we present a potential benchmarking protocol that provides a pathway to experimentally quantify the performance of a quantum device compared to our simulations. To illustrate this, we present the quantitative dispersion of the return probability by employing the noisy QFT subroutine. Our workflow was demonstrated on systems consisting of up to 128 qubits, which highlights the scalability of our methods. Even at this scale, by employing the QFT subroutine, we were able to provide empirical results thereby validating our power-dependent scaling predictions. 

We discuss ideas for future work that both generalize and specialize our current results. For example, our results so far are limited to specific noise features (i.e., noise described by an OU process). It will be of interest to consider noise extracted from quantum noise spectroscopy experiments. This can be done on a case-by-case basis to compare superconducting~\cite{Bylander2011}, trapped-ion~\cite{Wang2023}, nitrogen vacancy centers~\cite{Romach2015}, spin qubits in semiconductors~\cite{Chan2018} among other architectures. In addition, for now, we have only examined the QFT algorithm. Moving forward, this analysis can be performed with an enlarged test-suite of algorithms. It will be valuable to derive specialized results for variational as well as algorithms with provable precision guarantees. Away from computational algorithms, it is of interest to perform similar analyses with error correction, communication, and sensing protocols. 

\section{Acknowledgments}
A.G. and E.D. are supported by the U.S. Department of Energy, Office of Science, Advanced Scientific Research Program, Early Career Award under contract number ERKJ420. G.Q. acknowledges support from the U.S. Department of Energy, Office of Science, Office of Advanced Scientific Computing Research, Accelerated Research in Quantum Computing under Award Number DE-SC0025509 and Office of Science, Office of Fusion Energy Sciences, under Award Number DE-SC0025203. This research used resources of the Compute and Data Environment for Science (CADES) at the Oak Ridge National Laboratory, which is supported by the Office of Science of the U.S. Department of Energy under Contract No. DE-AC05-00OR22725.

\appendix

\section{SchWARMA models and Riemannian manifolds\label{app:riemannian_schwarma}}
To describe the SchWARMA models, we borrow a few concepts from differential geometry involving Riemannian manifolds. Intuitively, a manifold can be thought of as a space that is locally Euclidean. If equipped with a Riemannian metric,  it is a Riemannian manifold. In the current context, we consider the Stiefel manifold, $\mathcal{M}$ defined as a Riemannian manifold of isometries, i.e., a space of matrices, $\{V\}$ equipped with an inner product metric satisfying the constraints $V^{\dag}V = \mathds{1}$. Every element on the manifold $\mathcal{M}$, say $V$, is equipped with a tangent space, $\mathcal{T}_{V}$. To map elements of a particular tangent space back onto the manifold, we employ retraction functions, $\mathcal{R}$. Given an isometry $V$, to construct a corresponding noisy unitary associated with $V$, SchWARMA models explore the tangent space of $V$ further retracting the element on the tangent space back onto the manifold. 

\section{Relation to Prior Scaling}
\label{app:infidelity_to_delta}
In the main text we began by discussing some previous scaling results for the unitary operator error $|| U - \overline{\tilde{U}}||_{2} = \delta$. It will also be useful to recall that $\overline{\tilde{U}} = \mathbb{E}_l[\tilde{U}_l]$. Here, we relate $\delta$ to quantities computed in this work. 

On the other hand, the fidelity was \begin{eqnarray}
    \mathcal{F}_\eta&=& \frac{1}{n_{t}} \sum_{l=1}^{n_{t}}  |\langle  \eta  |U^\dagger \tilde{U}_l | \eta \rangle|^2 \\
    &=& \frac{1}{n_{t}} \sum_{l=1}^{n_{t}}  |a_l|^2 \\
    &=& \mathbb{E}_l[|a_l|^2]
\end{eqnarray}
where we have now explicitly written this quantity in terms of the expected value of the amplitudes $a_l$.

The intermediate quantity relating these two objects is $\bar{a} = \mathbb{E}_l[a_l] = \frac{1}{N}\sum_l \bra{\eta} U^\dagger \tilde{U}_l \ket{\eta} = \bra{\eta} U^\dagger \bar{\tilde{U}}_l \ket{\eta}$. To see this, examine $1-\mathbb{E}_l[a_l]$. That is, 
\begin{eqnarray}
    1 - \bar{a} &=& 1-\bra{\eta} U^\dagger \bar{\tilde{U}}_l \ket{\eta} \\
    &=& \bra{\eta} U^\dagger U - U^\dagger \bar{\tilde{U}}_l \ket{\eta} \\
    &=& \bra{\eta} U^\dagger (U - \bar{\tilde{U}}) \ket{\eta}.
\end{eqnarray}
Denoting $x= U \ket{\eta}$ and $y=(U - \bar{\tilde{U}}) \ket{\eta}$, we bound this quantity with Cauchy-Schwartz inequality to see that 
\begin{eqnarray}
    |1 - \bar{a}|&=& |x \cdot y| \leq ||x||_2 ||y||_2 \\
    &=&||(U - \bar{\tilde{U}}) \ket{\eta}||_2 \\
    &\leq& ||U - \bar{\tilde{U}}||_2 = \delta 
\end{eqnarray}
where we have used $||x||_2=1$. The second inequality is due to the fact that the spectral norm upper bounds the norm of $y$ (middle line). This then implies that $|\bar{a}|\geq 1-\delta$. 

Lastly, by the non-negativity of the variance, $ \mathbb{E}_l[|a_l|^2] \geq |\mathbb{E}_l[a_l]|^2$, tells us that $\mathcal{F}_\eta \geq |\bar{a}|^2$. In turn, means that 
\begin{eqnarray}
\mathcal{I} &=& 1-\mathcal{\mathcal{F}} \\ &\leq& 1- (1-\delta)^2 \\
&=& 2 \delta - \delta^2.
\end{eqnarray}

\bibliography{arxiv_submission/bib}
\end{document}